\documentclass[twocolumn]{revtex4}

\newcommand{\be}{\begin{eqnarray}}
\newcommand{\ee}{\end{eqnarray}}
\newcommand{\ket}[1]{\ensuremath{\left| {#1} \right>}}

\usepackage[latin1]{inputenc}
\usepackage{graphicx}
\usepackage{amsmath,amssymb}
\usepackage{hyperref}
\usepackage{latexsym}
\usepackage[squaren,thickspace,thickqspace]{SIunits}
\usepackage{multirow}
\usepackage{textcomp}
\usepackage{pdflscape}
\usepackage{rotating}
\usepackage{pbox}
\DeclareMathOperator\erf{erf}

\begin{document}
\title{Parallel transport quantum logic gates with trapped ions}

\author{Ludwig E. de Clercq}
\author{Hsiang-Yu Lo\footnote{These authors contributed equally to this work}}\email{hylo@phys.ethz.ch}
\author{Matteo Marinelli$^\ast$}
\author{David Nadlinger}
\author{Robin Oswald}
\author{Vlad Negnevitsky}
\author{Daniel Kienzler}
\author{Ben Keitch\footnote{Present address: Department of Engineering Science, University of Oxford, Parks Road, Oxford OX1 3PJ, UK}}
\author{Jonathan P. Home}

\address{Institute for Quantum Electronics, ETH Z\"urich, Otto-Stern-Weg 1, 8093 Z\"urich, Switzerland}


\begin{abstract}
We demonstrate single-qubit operations by transporting a beryllium ion with a controlled velocity through a stationary laser beam. We use these to perform coherent sequences of quantum operations, and to perform parallel quantum logic gates on two ions in different processing zones of a multiplexed ion trap chip using a single recycled laser beam. For the latter, we demonstrate individually addressed single-qubit gates by local control of the speed of each ion. The fidelities we observe are consistent with operations performed using standard methods involving static ions and pulsed laser fields. This work therefore provides a path to scalable ion trap quantum computing with reduced requirements on the optical control complexity.
\end{abstract}

\pacs{pacs}
\maketitle

A quantum computer will require large numbers of operations to be performed in parallel \cite{02Kielpinski, 05Knill, 03Steane, 07Steane} which should be achieved with as little technical overhead as possible. For trapped ions, operations are commonly performed using optical and microwave fields. Optical fields provide relatively simple ion addressing and increased field gradients, allowing for fast and high-fidelity single and multi-qubit operations \cite{08Benhelm, 14Ballance}. However, laser beam control typically involves the use of bulky acousto-optic devices to switch fields on and off, with pulse timing dependent on the intensity of the light reaching the ion. In order to implement parallel operations on multiple ions across a large-scale processor, the use of a single modulator would require that the intensities at each ion differ by less than $1\%$ for an error-per-gate of $10^{-4}$. This appears challenging to achieve across a large-scale device.  An alternative approach is to make the ion experience a pulse of light by transporting it through a static laser beam \cite{07Leibfried}. This has the advantage that the light need not be actively controlled, and the necessary control of the gate parameters can be integrated into the control of analog voltages applied to the trap electrodes. Since the control of the ion transport is local, different intensities of the light at different regions can be tolerated with no intrinsic loss in gate fidelity, by moving the respective ions at different speeds. Shuttling of ions is a key ingredient in one of the most promising schemes for scaling up trapped-ion quantum computing \cite{98Wineland2, 00Kielpinski, 09Home}.

In this Letter, we demonstrate trapped-ion quantum gates by transporting ions through static laser beams. We apply time-varying electric potentials to 30 electrodes in a multi-zone segmented ion trap, continuously controlling the velocity of the potential well confining the ion. We use this to demonstrate single-qubit Rabi oscillations, a Ramsey sequence, and the ability to perform individually-addressed gates in parallel at different locations of the processor using a single recycled laser beam. We find the fidelities to be comparable to the performance of standard static gates in our system.

\begin{figure}[ht!]
\includegraphics[width=0.95\columnwidth]{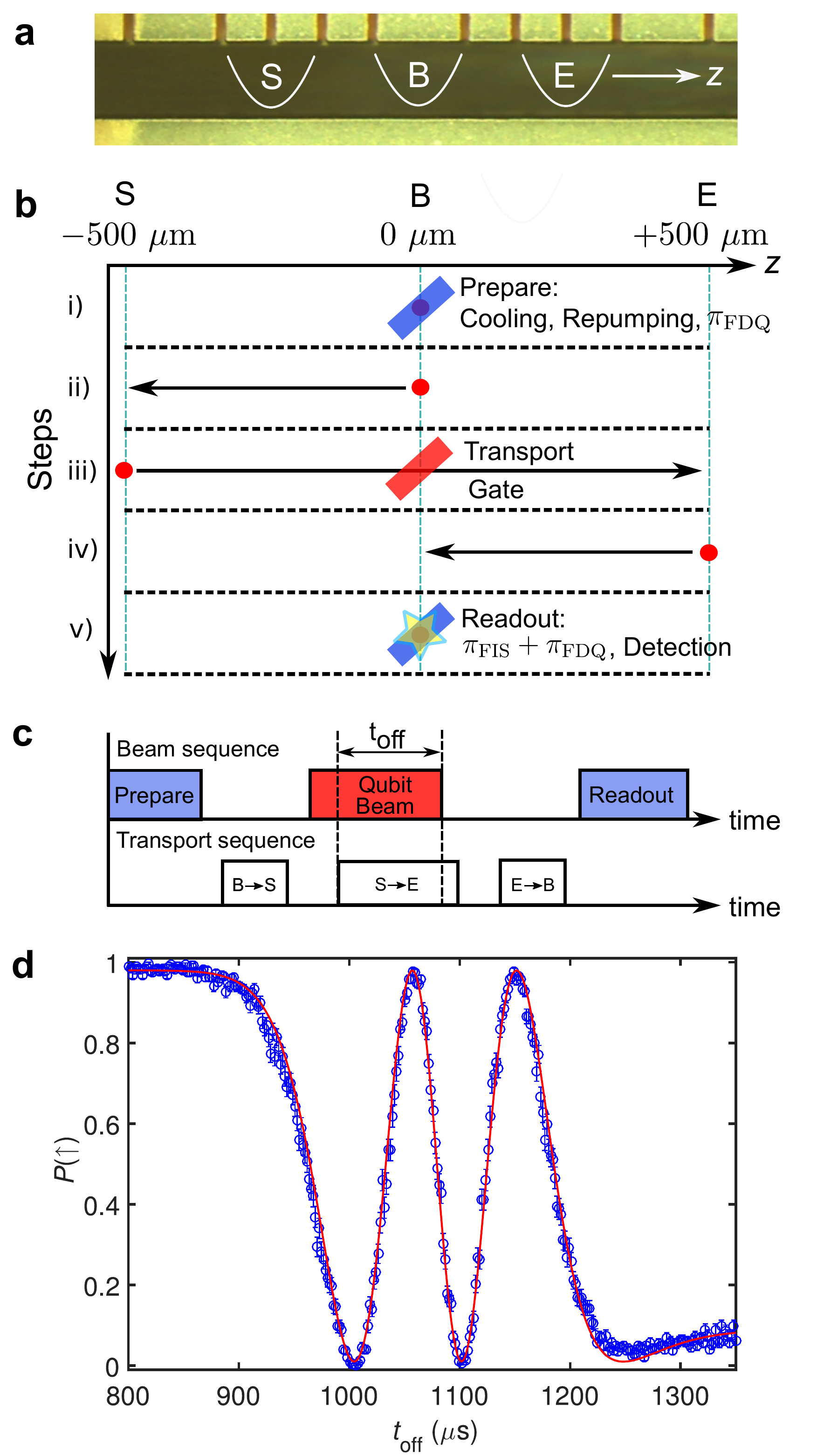}
\caption{ \textbf{a.} Image of the different trap zones along the transport axis $z$. \textbf{b.} The Rabi oscillation experiment is carried out using operations occurring in numerical order from i) to v). The use of the two coherent transfer pulses (labelled as $\pi_{\mathrm{FDQ}}$ and $\pi_{\mathrm{FIS}}$) is described in the text. \textbf{c.} The pulse sequence used to probe Rabi oscillations. The qubit beam is rapidly switched off at a time $t_{\rm off}$ during the transport in order to probe the dynamics of the spin. \textbf{d.} Rabi oscillations as a function of $t_{\rm off}$ for an ion transported through a two-frequency Raman laser beam pair. The fitted curve follows the theoretical prediction from Eq. (\ref{eq:Pup}), which assumes a Gaussian beam profile. Each data point represents the average of 350 repeats of the experimental sequence. Here and in the rest of the paper, error bars are estimated assuming quantum projection noise.}
	\label{fig:transportRabi}
\end{figure}

We work with qubits stored in the hyperfine levels of the ${^2}S_{1/2}$ ground state of beryllium ions. Working at a magnetic field of 11.945~mT, the $\ket{F=1,m_F=1}\leftrightarrow\ket{F=2,m_F=0}$ transition energy has zero first-order dependence on the magnetic field, making this an excellent memory qubit, which is robust against spatial magnetic field variations during transport \cite{05Langer}. Each experimental run starts with cooling all motional modes of the ion close to the Doppler limit. Preparing the ion in $\ket{F=1,m_F=1}$ state is achieved by optical pumping to $\ket{F=2,m_F=2}$ followed by a resonant stimulated Raman transition pulse (labelled as $\pi_{\mathrm{FDQ}}$ in the figures) which coherently transfers the population to $\ket{F=1,m_F=1}$ (referred to as $\ket{\uparrow}$ in the text). All the coherent operations used in this experiment are performed using co-propagating Raman beams, using a laser red-detuned by $\approx$230~GHz from the ${^2}S_{1/2}\leftrightarrow {^2}P_{1/2}$ transition \cite{14Lo}.
For the transport gates, we shuttle the ion between two different regions of the trap, passing through the laser beams during the transport.
Following the quantum logic gate, we read out the qubit state by first performing two coherent Raman transfer pulses, the first of which (labelled as $\pi_{\mathrm{FIS}}$ in the figures) shelves the population from state $\ket{F=2,m_F=0}$ to $\ket{F=1,m_F=-1}$ for improving the readout fidelity \cite{09Home, ThesisLo}, while the second transfers the population from $\ket{F=1,m_F=1}$ to $\ket{F=2,m_F=2}$. Subsequent readout of the quantum state is performed by state-dependent fluorescence applied to the ${^2}S_{1/2}\ket{F=2,m_F=2}\leftrightarrow{^2}P_{3/2}\ket{F'=3,m_{F'}=3}$ cycling transition \cite{98Wineland2,95Monroe}.
We transport ions by applying time-dependent potentials to the trap electrodes, which are generated using a home-built 16-channel arbitrary waveform generator (AWG). We calculate the required waveforms using a constrained optimization method, allowing control of  multiple potential wells with constrained position, curvature and depth simultaneously.

For initial experiments with a single ion, we prepare and read out the state in zone B (Fig. \ref{fig:transportRabi}a). The transport gate is implemented by first transporting the ion to zone S, which is outside the intensity profile of the laser beams used to perform the logic gate (Figs. \ref{fig:transportRabi}b and c). We then turn on the gate Raman beam pair, which crosses the trap axis near the center of zone B, with the beams making an angle of 45 degrees to the trap axis. We subsequently transport the ion from zone S to zone E, passing through zone B. All transport sequences are adiabatic, and thus the ion position closely follows the potential well minimum \cite{12Bowler, 12Walther}. We aim to achieve a constant velocity during transport, but have observed in separate experiments with calcium ions that this is not achieved due to an interplay between the outputs of our AWG and passive noise filters which connect these to the ion trap \cite{15deClercqDoppler}. Nevertheless, by tuning the velocity we are able to control the gate operation performed on the ions.

\begin{figure}
	\centering
	\includegraphics[width=0.95\columnwidth]{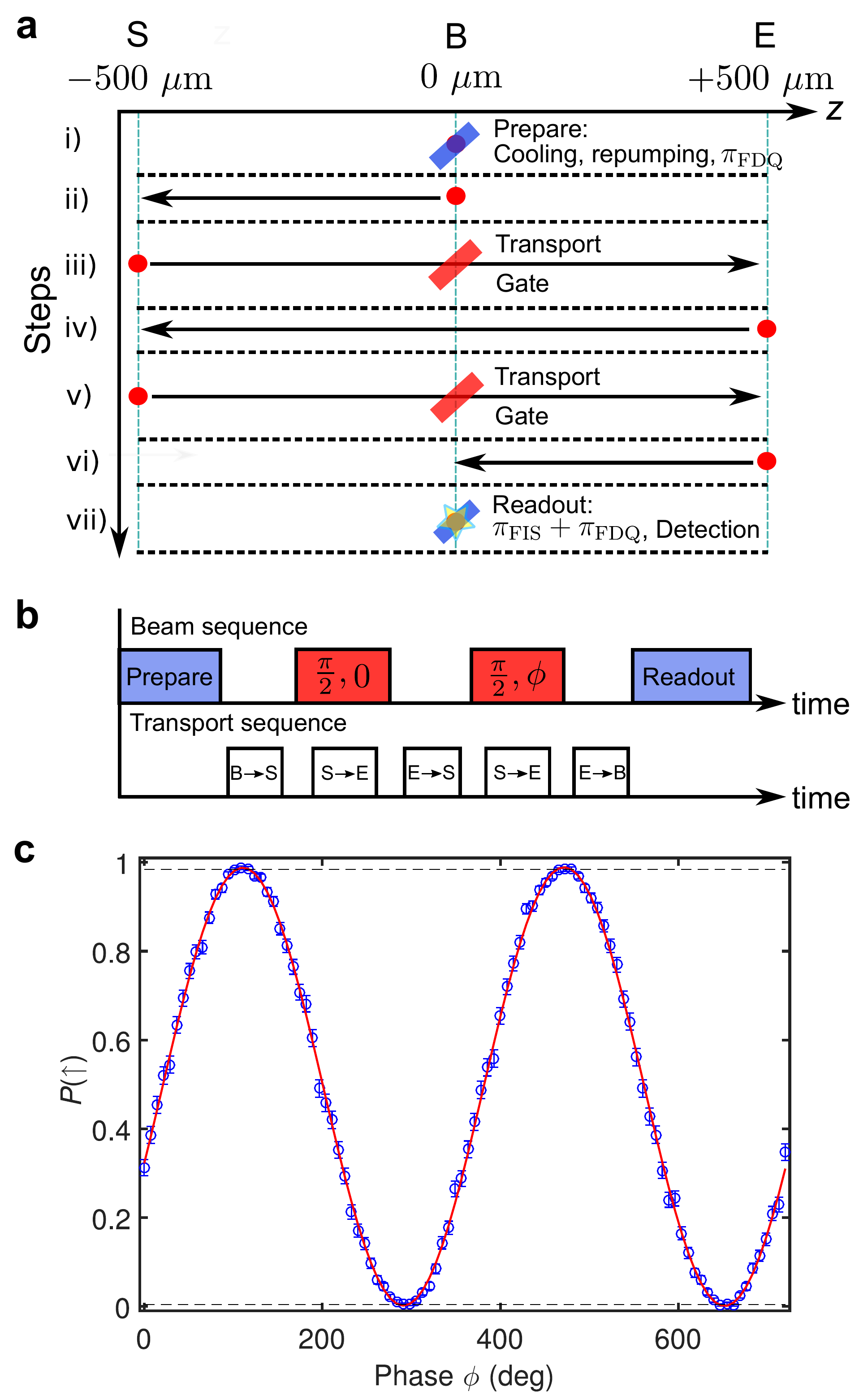}
	\caption{\textbf{a.} The Ramsey sequence involves steps i) to vii). \textbf{b.} The pulse sequence used for the Ramsey experiment. We concatenate two transport $\pi/2$ gates in which the phase of the laser during the second gate is shifted by $\phi$ relative to the first. In this case, the qubit beams remain on until the ion has passed through to zone E. \textbf{c.} Experimental results for a Ramsey experiment. The phase $\phi$ is scanned using a shift in the relative phase of the two Raman beams applied with an acousto-optic modulator. The contrast of the oscillations obtained from the fit is 0.985 $\pm$ 0.004. For comparison, the plotted horizontal dashed lines indicate the maximum and minimum of Rabi oscillations performed using standard static-ion gates. Each data point is the average of 600 repeats of the experimental sequence.}
	\label{fig:transportRamsey}
\end{figure}

The Rabi oscillations shown in Fig. \ref{fig:transportRabi}d are obtained by recording the population of $\ket{\uparrow}$ as a function of the transit time through the beam. This is achieved by switching off the laser beams rapidly during the transport. For an ion transported at a constant velocity $v$ through a Gaussian beam profile with waist $w_0$ and a peak Raman transition Rabi frequency $\Omega_0$, the probability to find the ion in the $\ket{\uparrow}$ state as a function of time is given by \cite{07Leibfried}
\be
P(\uparrow, t) = \cos^2(\zeta(t)/2), \ \zeta(t) = \frac{\Omega_0}{\chi} \sqrt{\pi}\  f(t, t_0)
\label{eq:Pup}
\ee
with $f(t, t_0) = \erf(\chi t_0) - \erf(\chi(t_0 - t))$ and $\chi = v/\sqrt{2} w_0$. The parameter $t_0$ accounts for the time at which the ion transits the beam. We fit the data with a form similar to Eq. (\ref{eq:Pup}) but including offset and amplitude parameters which account for imperfections in state preparation, readout and the Rabi oscillation itself. We obtain $\Omega_0/(2 \pi) = 5.669 \pm 0.001$~kHz and $\chi=7753\pm 23$~s$^{-1}$. Using the relationship between  velocity $v$ and beam waist $\omega_0$ given in Eq. (\ref{eq:Pup}), we use an independently measured beam waist to deduce a transport velocity of $0.62\pm 0.05$~m{\,}s$^{-1}$.  To further improve the quality of the fit, Eq. (\ref{eq:Pup}) could be modified to take into account changes in the ion velocity during transport, and the deviations of the intensity of the laser beams from an ideal Gaussian \cite{15deClercqDoppler}.

Having verified the dynamics, we calibrate the laser intensity and the ion velocity in order to apply a single qubit rotation $\hat R(\pi/2)$, where $\hat R(\theta) \equiv \cos(\theta/2) \hat I - i \sin(\theta/2) \hat \sigma_x$. This requires a velocity of $\sim$7~m{\,}s$^{-1}$. Using two sequential transports (Figs. \ref{fig:transportRamsey}a and b), we perform a sequence of two gates on the ion, implementing a Ramsey experiment. Both gates produce the rotation $\hat R(\pi/2)$, however we shift the relative phase of the two Raman beams before applying the second transport gate. A scan of this phase is shown in Fig. \ref{fig:transportRamsey}c, which we fit with a sinusoid to obtain an amplitude of 0.985 $\pm$ 0.004. This is consistent with the amplitude we observe for Rabi oscillations performed with a static ion and pulsed laser fields. The second Ramsey pulse could also have been performed during the transport occuring at step iv) in Fig. \ref{fig:transportRamsey}a, which we think would have a negligible impact on the fidelity. We chose to apply the gate instead in step v) for experimental simplicity.

\begin{figure}
\centering
	\includegraphics[width=0.95\columnwidth]{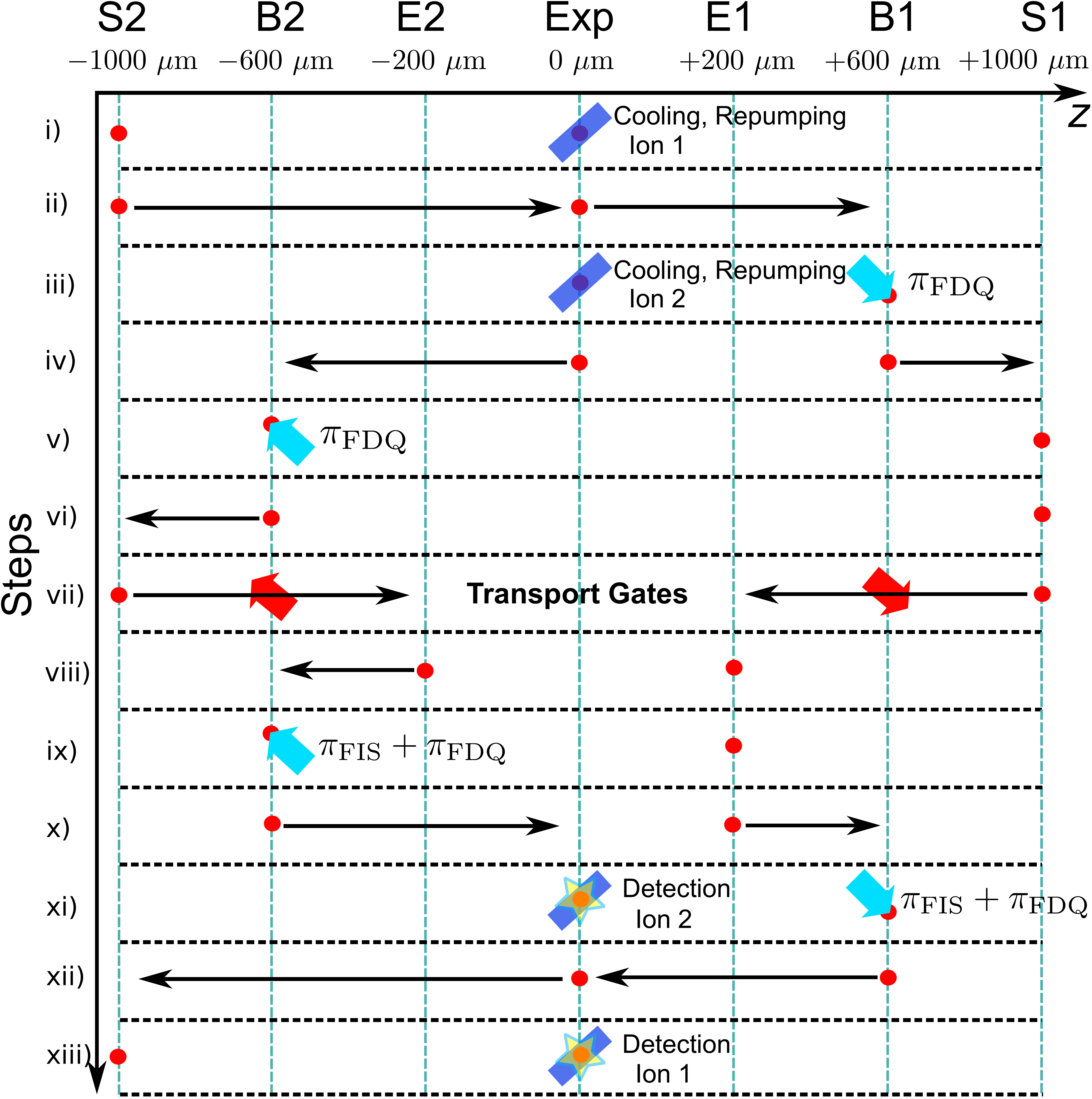}
	\caption{For work on parallel gates, the steps i) through to xiii) are carried out sequentially. Optical pumping to $\ket{F=2,m_F=2}$ and state detection are performed sequentially in the experimental zone using pulsed operations, which are also used for coherent transfer pulses ($\pi_{\mathrm{FDQ}}$ and $\pi_{\mathrm{FIS}}$) in the separated zones. The transport gates are applied simultaneously, by passing the ions through qubit beams located at $\pm$~600~$\mu$m from the trap center.}
	\label{fig:trapsequence_multi}
\end{figure}

\begin{figure}
\centering
	\includegraphics[width=\columnwidth]{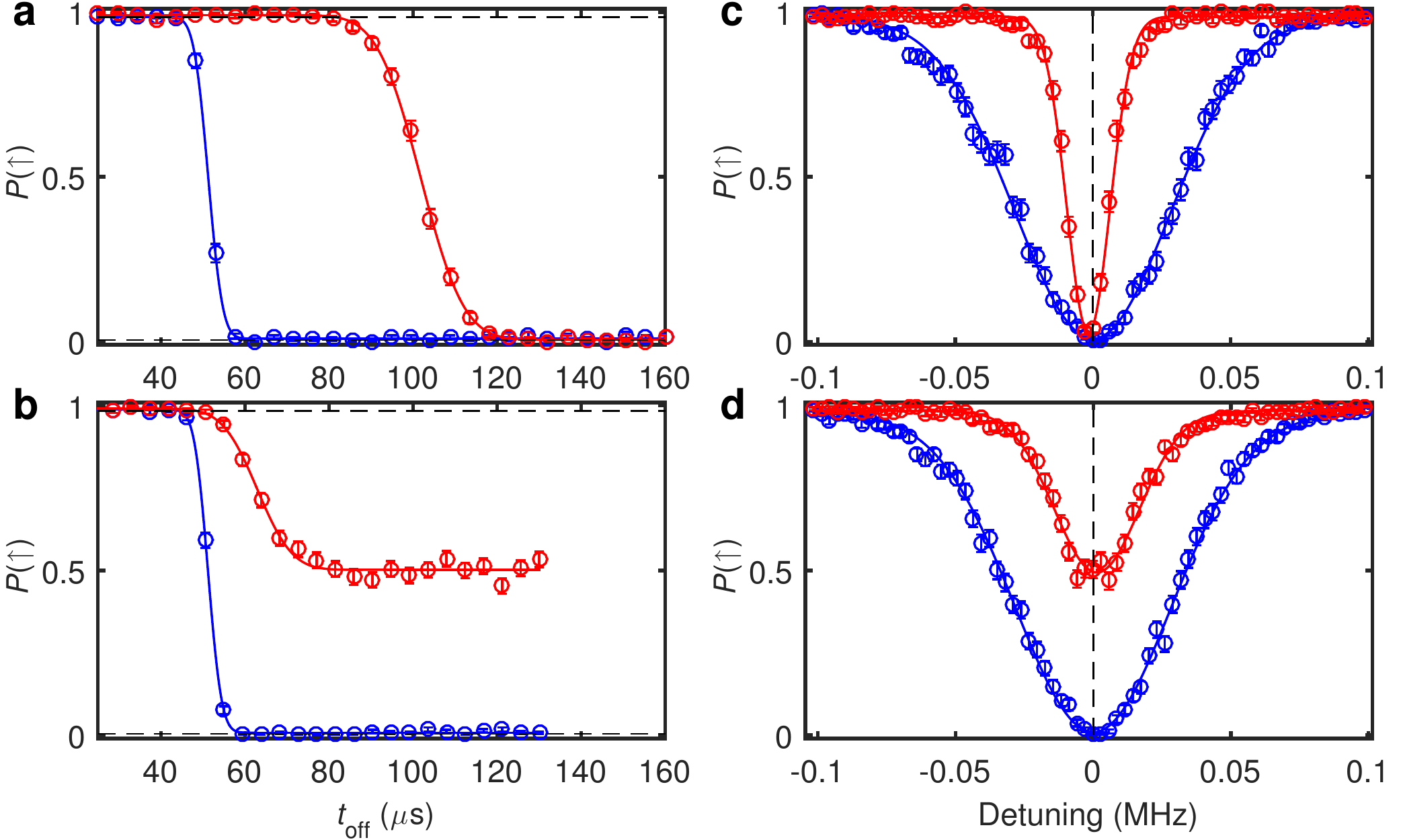}
	\caption{\textbf{a}, \textbf{b.} Time scans for the simultaneous rotation of Ion 1 (blue) and Ion 2 (red) with two different velocities for Ion 2 resulting in a rotation $\hat R(\pi)$ in \textbf{a} and $\hat R(\pi/2)$ in \textbf{b} respectively. The dashed horizontal lines indicate the maximum and minimum of Rabi oscillations performed using static-ion gates. Corresponding scans of the difference frequency of the Raman lasers with $t_{\mathrm{off}} > 150~\mu\mathrm{s}$ are shown in \textbf{c}, \textbf{d.}, which we fit with Gaussian functions. Vertical dashed lines indicate the centre of the fit to the blue data points. Each data point is the average of 250 repeats of the experimental sequence.}
	\label{fig:Parallel}
\end{figure}

Figure \ref{fig:trapsequence_multi} shows a second set of experiments in which we work with two ions in parallel. To do this, we shift the Raman beams to pass through the trap in zone B1 (600 \textmu m away from the trap center), and then reflect these laser fields back through the vacuum setup to pass through a focus close to the axis of the trap in zone B2. The experimental sequence of operations is shown in Fig. \ref{fig:trapsequence_multi}. Preparation and measurement steps are performed sequentially for the two ions using the same methods as in the single ion experiments. For the transport gate operation, Ion 1 is transported from zone S1 to zone E1 passing through the beams centered at B1, while Ion 2 is transported from zone S2 to zone E2 through the beams centered at B2. Readout is then performed sequentially. Data for two different transport velocities of Ion 2 are shown in Fig. \ref{fig:Parallel}, resulting in a $\hat R(\pi)$ or $\hat R(\pi/2)$ pulse being implemented. The transport velocity of Ion 1 is calibrated to be 10.7~m{\,}s$^{-1}$ to produce a $\hat R(\pi)$ rotation and remains the same for both data sets. For Ion 2, the velocity is calibrated to either produce $\hat R(\pi)$ (requiring $v\approx 4.7$~m{\,}s$^{-1}$) or $\hat R(\pi/2)$ (requiring $v\approx 8.9$~m{\,}s$^{-1}$). We fit the data with a function of the form $\frac{1}{2}(a+b \erf (s(t-t_c)))$, where $a$ is the offset, $b$ is the amplitude, $s$ is the steepness and $t_c$ is the time shift of the error function. We obtain $b = 0.978\pm 0.002$ for the $\hat R(\pi)$ rotations for Ion 1 (blue traces in Figs. \ref{fig:Parallel}a and \ref{fig:Parallel}b). For Ion 2, we obtain $b = 0.986 \pm 0.002$ for the $\hat R(\pi)$ rotation and a final population of $0.502\pm 0.004$ for the $\hat R(\pi/2)$ rotation (red traces in Figs. \ref{fig:Parallel}a and \ref{fig:Parallel}b). These results demonstrate the individual addressing capability in the transport gates while simultaneously performing operations with a single pair of Raman laser beams.

In the experiments performed above, the Raman laser beams are reflected back into the trap after passing out of the vacuum system. This produces two significant effects. Firstly there is a loss in optical power due to imperfect transmission of the vacuum windows and other optical components. Secondly, we observe an increase of the beam size in B2 relative to B1. Experimentally we measure the waist diameter of the beam in B1 to be 73 $\pm$ 2~\textmu m and a peak Rabi frequency of  $2\pi\times(47.9 \pm 0.5$~kHz$)$ while the retro-reflected beams in B2 have a waist diameter of 160 $\pm$ 4~\textmu m and a peak Rabi frequency of $2\pi\times(16.1 \pm 0.5$~kHz$)$. The significant difference in intensities leads to a differential AC Stark shift between the two ions which we measure using static Rabi oscillations to be $1.3 \pm 0.2$~kHz, agreeing with the theoretical calculation \cite{03Wineland}. By tuning the difference frequency of the Raman light fields close to the resonance of the less intense beams, we obtain the average operation fidelity in zone B1 (the location of the more intense beams) due to this residual detuning being $F = 0.9998$ for a $\pi/2$ rotation \cite{07Pedersen}. This effect could be reduced by better equalization of the beam intensity, both through improved focussing and by the use of in-vacuum mirrors.

The use of co-propagating beams has the significant advantage that the difference wave vector is close to zero, which minimizes the first-order Doppler shift. In our current setup the two Raman beams are generated by passing each through a separate optical path, and combining them on a beamsplitter. We notice that even small misalignments lead to a differential Doppler shift (a relative angle of 10$^{-3}$ degrees gives a Doppler shift of around 400~Hz for a velocity of 10 m{\,}s$^{-1}$). We minimize this by performing two frequency scans, where the transport direction of one of two ions is reversed between each scan, and adjust the beam alignment to null out the corresponding change in the resonance frequency. However residual effects remain, which we think causes the residual skew of the data in Fig. \ref{fig:Parallel}. The use of common optical fibers for both Raman beams would alleviate this problem \cite{14Gebert, 14Colombe}.

The contrast of operations performed using transport gates is the same as for our standard pulsed static gates to within experimental error, which indicates that the transport itself does not produce a detectable additional error. In our current setup, the population transfer from states ${^2}{S}_{1/2}\ket{F=2,m_F=2}$ to ${^2}{S}_{1/2}\ket{F=1,m_F=1}$ and vice versa produce the biggest source of error. From calculations using our experimental detection time, we estimate the static-ion transfer pulses to have a fidelity of around 99.2\%. Better characterization of the transport gates themselves could be performed using randomized benchmarking \cite{08Knill, 12Gaebler}.

Transport-based quantum logic gates provide significant advantages for scaling up ion trap quantum processors. Waveforms applied to the trap electrodes could in the future be performed using in-vacuum electronics, and possibly integrated into the trap structure itself through the use of CMOS or other well-established technologies \cite{14Mehta}. This would shift the control challenge from bulky optical setups to the well-developed field of micro-electronics, which has proven scalability. The use of recycled laser beams to address ions in many parts of a trap array would optimize the use of laser power, which is a precious resource since lasers are expensive and bulky.

We thank Joseba Alonso and Christa Fl{\"u}hmann for contributions to the experimental apparatus, and Florian Leupold for comments on the manuscript. We acknowledge support from the Swiss National Science Foundation under grant no. $200021\_134776$ and 200020\_153430, ETH Research Grant under grant no. ETH-18 12-2, and from the Swiss National Science Foundation through the National Centre of Competence in Research for Quantum Science and Technology (QSIT).\\


\begin{thebibliography}{10}


\bibitem{02Kielpinski}
D.~Kielpinski, C.~Monroe, and D.~Wineland.
\newblock Architecture for a large-scale ion-trap quantum computer.
\newblock {\em Nature}, 417:709--711, 2002.

\bibitem{05Knill}
E.~Knill.
\newblock Quantum computation with realistically noisy devices.
\newblock {\em Nature}, 434:39--44, 2005.

\bibitem{03Steane}
A.~M. Steane.
\newblock Overhead and noise threshold of fault-tolerant quantum error
  correction.
\newblock {\em Phys. Rev. A}, 68:042322, 2003.

\bibitem{07Steane}
A.~Steane.
\newblock How to build a 300 bit, 1 Giga-operation quantum computer.
\newblock {\em Quant. Inf. and Comp.}, 7:171, 2007.

\bibitem{08Benhelm}
J.~Benhelm, G.~Kirchmair, C.~F. Roos, and R.~Blatt.
\newblock Towards fault-tolerant quantum computing with trapped ions.
\newblock {\em Nature Physics}, 4:463--466, 2008.

\bibitem{14Ballance}
C.~J. Ballance, T.~P. Harty, N.~M. Linke, and D.~M. Lucas.
\newblock High-fidelity two-qubit quantum logic gates using trapped calcium-43
  ions.
\newblock {\em arXiv preprint arXiv:1406.5473}, 2014.

\bibitem{07Leibfried}
D.~Leibfried, E.~Knill, C.~Ospelkaus, and D.~J. Wineland.
\newblock Transport quantum logic gates for trapped ions.
\newblock {\em Phys. Rev. A}, 76:032324, 2007.

\bibitem{98Wineland2}
D.~J. Wineland, C.~Monroe, W.~M. Itano, D.~Leibfried, B.~E. King, and D.~M.
  Meekhof.
\newblock Experimental issues in coherent quantum-state manipulation of trapped
  atomic ions.
\newblock {\em J. Res. Natl. Inst. Stand. Technol.}, 103:259--328, 1998.

\bibitem{00Kielpinski}
D.~Kielpinski, B.~E. King, C.~J. Myatt, C.~A. Sackett, Q.~A. Turchette, W.~M.
  Itano, C.~Monroe, and D.~J. Wineland.
\newblock Sympathetic cooling of trapped ions for quantum logic.
\newblock {\em Phys. Rev. A}, 61:032310, 2000.

\bibitem{09Home}
J.~P. Home, D.~Hanneke, J.~D. Jost, J.~M. Amini, D.~Leibfried, and D.~J.
  Wineland.
\newblock Complete methods set for scalable ion trap quantum information
  processing.
\newblock {\em Science}, 325:1227, 2009.

\bibitem{05Langer}
C.~Langer, R.~Ozeri, J.D. Jost, J.~Chiaverini, B.L. DeMarco, A.~Ben-Kish, R.B.
  Blakestad, J.~Britton, D.~Hume, W.M. Itano, D.~Leibfried, R.~Reichle,
  T.~Rosenband, T.~Schaetz, P.O. Schmidt, and D.J. Wineland.
\newblock Long-lived qubit memory using atomic ions.
\newblock {\em Phys. Rev. Lett.}, 95:060502, 2005.

\bibitem{14Lo}
H.-Y. Lo, J. Alonso, D. Kienzler, B.~C. Keitch, L. de~Clercq,
  V. Negnevitsky, and J.~P. Home.
\newblock All solid-state continuous-wave laser systems for ionization, cooling
  and quantum state manipulation of beryllium ions.
\newblock {\em Appl. Phys. B}, 114:17--25, 2014.

\bibitem{ThesisLo}
H.-Y. Lo.
\newblock Creation of Squeezed Schr\"odinger's Cat States in a Mixed-Species Ion Trap.
\newblock PhD Thesis, ETH Z{\"u}rich, Switzerland, 2015.

\bibitem{95Monroe}
C. Monroe, D. M. Meekhof, B. E. King, S. R. Jefferts, W. M. Itano, D. J. Wineland and P. Gould.
\newblock Resolved-sideband {R}aman cooling of a bound atom to the {3D} zero-point energy.
\newblock {\em Phys. Rev. Lett.}, 75:4011, 1995.

\bibitem{12Bowler}
R.~Bowler, J.~Gaebler, Y.~Lin, T.~R. Tan, D.~Hanneke, J.~D. Jost, J.~P. Home,
  D.~Leibfried, and D.~J. Wineland.
\newblock Coherent diabatic ion transport and separation in a multizone trap
  array.
\newblock {\em Phys. Rev. Lett.}, 109:080502, 2012.

\bibitem{12Walther}
A.~Walther, F.~Ziesel, T.~Ruster, S.~T. Dawkins, K.~Ott, M.~Hettrich,
  K.~Singer, F.~Schmidt-Kaler, and U.~Poschinger.
\newblock Controlling fast transport of cold trapped ions.
\newblock {\em Phys. Rev. Lett.}, 109:080501, 2012.

\bibitem{15deClercqDoppler}
L.~E. de~Clercq, R.~Oswald, C.~Fl{\"u}hmann, B.~Keitch, D.~Kienzler, H.-Y.
  Lo, M.~Marinelli, D.~Nadlinger, V.~Negnevitsky, and J.~P. Home.
\newblock Time-dependent Hamiltonian estimation for Doppler velocimetry of trapped ions.
\newblock {\em arXiv:1509.07083}, 2015.

\bibitem{03Wineland}
D.~J. Wineland, M.~Barrett, J.~Britton, J.~Chiaverini, B.~DeMarcoand W.~M.
  Itano, B.~Jelenkovic, C.~Langer, D.~Leibfried, V.~Meyer, and
  T.~Rosenbandand~T. Sch\"atz.
\newblock Quantum information processing with trapped ions.
\newblock {\em Phil. Trans. R. Soc. Lond. A}, 361:1349--1361, 2003.

\bibitem{07Pedersen}
L.~H. Pedersen, N.~M. Moeller, and K. M{\o}lmer.
\newblock Fidelity of quantum operations.
\newblock {\em Physics Letters A}, 367:47--51, 2007.

\bibitem{14Gebert}
F.~Gebert, M.~H. Frosz, T.~Weiss, Y.~Wan, A.~Ermolov, N.~Y. Joly, P.~O.
  Schmidt, and P.~St.~J. Russell.
\newblock Damage-free single-mode transmission of deep-uv light in hollow-core
  pcf.
\newblock {\em Opt. Express}, 22:15388--15396, 2014.

\bibitem{14Colombe}
Y. Colombe, D.~H. Slichter, A.~C. Wilson, D. Leibfried, and
  D.~J. Wineland.
\newblock Single-mode optical fiber for high-power, low-loss uv transmission.
\newblock {\em Opt. Express}, 22:19783--19793, 2014.

\bibitem{08Knill}
E.~Knill, D.~Leibfried, R.~Reichle, J.~Britton, R.~B. Blakestad, J.~D. Jost,
  C.~Langer, R.~Ozeri, S.~Seidelin, and D.~J. Wineland.
\newblock Randomized benchmarking of quantum gates.
\newblock {\em Phys. Rev. A}, 77:012307, 2008.

\bibitem{12Gaebler}
J.~P. Gaebler, A.~M. Meier, T.~R. Tan, R.~Bowler, Y.~Lin, D.~Hanneke, J.~D.
  Jost, J.~P. Home, E.~Knill, D.~Leibfried, and D.~J. Wineland.
\newblock Randomized benchmarking of multiqubit gates.
\newblock {\em Phys. Rev. Lett.}, 108:260503, 2012.

\bibitem{14Mehta}
K.~K. Mehta, A.~M. Eltony, C.~D. Bruzewicz, I.~L. Chuang, R.~J. Ram, J.~M.
  Sage, and J.~Chiaverini.
\newblock Ion traps fabricated in a CMOS foundry.
\newblock {\em Applied Physics Letters}, 105:044103, 2014.

\end{thebibliography}

\end{document}